\documentclass[useAMS,usenatbib]{tellus}
\usepackage{amsmath, graphicx, natbib}
\usepackage[hyperindex,breaklinks]{hyperref}

\begin{document}

\title[Joint estimation of the state]{Ensemble regional data assimilation using joint states}

\author[Y. Yoon et al.]{By Young-noh Yoon$^1$\thanks{Corresponding author.\hfil\break e-mail: mystyle@umd.edu}, Brian R. Hunt$^2$, Edward Ott$^3$ and Istvan Szunyogh$^4$,} \affiliation{$^{1}$Department of Physics, University of Maryland, College Park, MD, USA; $^{2}$Department of Mathematics and Institute for Physical Science and Technology, University of Maryland, College Park, MD, USA; $^{3}$Institute for Research in Electronics and Applied Physics, University of Maryland, College Park, MD, USA; $^{4}$Department of Atmospheric Sciences, Texas A\&M University, College Station, TX, USA}

\history{\today}

\maketitle

\begin{abstract}
We propose a data assimilation scheme that produces the analyses for a global and an embedded limited area model simultaneously, considering forecast information from both models. The purpose of the proposed approach is twofold. First, we expect that the global analysis will benefit from incorporation of information from the higher resolution limited area model. Second, our method is expected to produce a limited area analysis that is more strongly constrained by the large scale flow than a conventional limited area analysis. The proposed scheme minimizes a cost function in which the control variable is the joint state of the global and the limited area models. In addition, the cost function includes a constraint term that penalizes large differences between the global and the limited area state estimates. The proposed approach is tested by idealized experiments, using `toy' models introduced by Lorenz in 2005. The results of these experiments suggest that the proposed approach improves the global analysis within and near the limited area domain and the regional analysis near the lateral boundaries. These analysis improvements lead to forecast improvements in both the global and the limited area models.\\\\\\
\end{abstract}

\section{Introduction}

Assuming that we have a global model and a regional model of higher accuracy defined in a subregion inside the global region, we aim to produce a forecast which is better than the one from each model by using information from both models. We test two data assimilation methods. The first method is based on techniques most commonly used in current practice and has recently been tested in \citet{Merkova-et-al-11}. In this method, the global and the regional data assimilations are done separately, and the regional model receives information from the global model through the boundaries during the integration phase, but the global model does not receive information from the regional model. The second method, which we call the joint state method, is proposed in this paper. In this method, the global and regional data assimilations are coupled simultaneously using information contained in both the global and the regional forecast states, and the regional model receives information from the global model through the boundaries during the integration phase as in the separate analysis method. We use the Local Ensemble Transform Kalman Filter (LETKF) algorithm for data assimilation. This algorithm allows efficient implementation of the localization technique proposed by \citet{Ott-et-al-04}. In order to test our global/regional assimilation techniques we use numerical experiments based on simple atmospheric `toy' models proposed in \citet{Lorenz-05} in conjunction with simulated observations. We compare results of our joint state method and results of the separate analysis method. We find that better forecasts are produced by using the joint state method than by using the separate analysis method. We note that our proposed scheme would most likely be of potential interest for centers, where both global and limited area forecasts and analyses are prepared.

The organization of the paper is the following. Section~\ref{model} introduces the atmospheric toy models that we use. Section~\ref{analysis} describes the data assimilation schemes by the joint state method and by the separate analysis method. Section~\ref{integration} describes how the regional model is coupled to the global model at the boundaries of the subregion during the integration phases of forecast cycles. Section~\ref{results} compares the results of our joint state method to those of the separate analysis method. Section~\ref{conclusion} gives further discussion and summarizes our conclusions.

\section{True model, global model, and regional model}\label{model}

\citet{Lorenz-05} introduced three simple, spatially discrete, 1-dimensional models that have been proven to be useful for testing weather data assimilation methods. Here we will use Lorenz's model 2 (which shows smooth propagating waves) and the more refined Lorenz model 3 (which shows small scale activities on top of smooth waves). Lorenz model 3 is the following equation for the evolution of a scalar state variable $Z_n$ at spatial location $n$,
\begin{equation}
dZ_n/dt = [X,X]_{K,n} + b^2 [Y,Y]_{1,n} + c [Y,X]_{1,n} - X_n -b Y_n + F,\label{Z}
\end{equation}
where $n$ is an integer, $n=0,1,\ldots,N-1$, and $b$, $c$, and $F$ are parameters, and a periodic boundary condition is used ($Z_{N}=Z_{0}$). The convention of index counting starting from 0 is used throughout this paper. $N$-component vectors $X$ and $Y$ are defined as
\begin{align}
X_n &= \sum_{i=-I}^{I}{}'(\alpha - \beta |i|)Z_{n+i},\label{X}\\
Y_n &= Z_n - X_n,\label{Y}\\
\alpha &= (3I^2 + 3)/(2I^3+4I),\\
\beta &= (2I^2 + 1)/(I^4+2I^2),
\end{align}
where the prime notation on $\Sigma'$ signifies that the first and the last terms in the summation are divided by two, and $I$ is a parameter. The bracket of any two vectors $X$ and $Y$ is defined as
\begin{align}
[X,Y]_{K,n} = &\sum_{j=-J}^{J}{}' \sum_{i=-J}^{J}{}' ( -X_{n-2K-i} Y_{n-K-j}\nonumber\\
&+ X_{n-K+j-i} Y_{n+K+j} ) / K^2\label{bracket}
\end{align}
when $K$ is even, and $\Sigma'$ is replace by $\Sigma$ when $K$ is odd; $J=K/2$ when $K$ is even, and $J=(K-1)/2$ when $K$ is odd, where $K$ is a parameter.  Model 3 reduces to model 2 when $I=1$. In particular, for $I=1$, Eq.~(\ref{X}) yields $X_n=Z_n$, which by Eq.~(\ref{Y}) implies that $Y_n=0$. Thus, after changing notation, $n \rightarrow m$ and $Z_n \rightarrow Z_m$, we obtain
\begin{equation}
dZ_m/dt = [Z,Z]_{K,m} - Z_m + F,\label{model2}
\end{equation}
where $m$ is used to denote a point on the coarser grid of the global model.

We use Lorenz model 3 with parameter values $N = 960, K = 32, b = 10, c = 0.6, F = 15, I = 12$ to generate our simulated true dynamics, and Lorenz model 2 with $N = 240, K = 8, F = 15$ for the global model defined at every fourth grid point of the true model ($n=0,4,8,\ldots,956$). Thus the grid points for Eq.~(\ref{model2}) occur at $m=n/4$, where $n=0, 4, 8, \ldots, 956$. We assume that, between analyses, Eq.~(\ref{model2}) for $Z_m$ gives an approximation of the dynamical evolution of $Z_n(t)$ at the grid points $n=4m$. When referring to locations or lengths of regions, we use the coordinate system of the true model throughout this paper ($n=0,1,\ldots,959$). For the regional model, we define a subregion extending from $n=n_{0}=240$ to $n=n_{1}=720$ grid, and use Lorenz model 3 with the same parameter values as the true model. In order to integrate this regional model, we must evaluate the bracket quantities on the right hand side of Eq.~(\ref{Z}) defined by Eq.~(\ref{bracket}). For $n$ too close to $n_0$ ($n_1$) this involves $X$, $Y$, and $Z$ values at grid points outside the subregion, $n<n_0$ ($n>n_1$). Also, from Eq.~(\ref{X}), $X_n$ in the regional model (and hence also $Y_n$) depends on $Z_{n'}$ values in $n'<n_0$ ($n'>n_1$) if $n$ is within a distance $I$ of $n_0$ ($n_1$). To evaluate these quantities, we use estimates of the required values of $Z_{n'}$ obtained from interpolation of the global values $Z_m$ onto the $n$-grid. These interpolations essentially play the role of boundary conditions for the regional model.

\section{Data assimilation}\label{analysis}

We selected 15 evenly spaced observation points starting from $n=0$ ($n=0,64,128,\ldots,896$). Notice that all the observation points are at grid points defined in the global model. We construct simulated observations by adding random noise drawn from independent Gaussian distributions of standard deviation 1 to the true state values at the observation points.

We compare two data assimilation methods. The first method does data assimilation for the global model and the regional model separately, while the second method, which we call the joint state method, forms a combined state from the global model and the regional model and does data assimilation on the combined state. The intuition motivating our second method is that we expect the global and the regional estimates will both benefit from information exchange between them. We use LETKF for both methods. See \citet{Hunt-et-al-07} for an explanation of LETKF.

For the separate analysis method, we use LETKF without much modification. For the global analysis, at each grid point $n=4m$ defined in the global model, we define a local patch $[n-s,n+s]$ of size $2s+1$ with $s=40$, use the Ensemble Transform Kalman Filter (ETKF) to obtain an analysis for the $(2s+1)$ state values in each patch. This yields local patch analyses for each ensemble member. As done by others \citep[e.g.,][]{Hunt-et-al-07}, we then use these patch analyses to form the global analysis states for each ensemble member by defining the value of the global ensemble field at each point $m=n/4$ to be the analysis state value of that ensemble member in the center of patch $n=4m$. For the regional analysis, at each grid point $n$ defined in the regional model, we define a local patch, limiting the size near the two boundaries of the subregion so that the local patch is defined only inside the subregion, use ETKF, and take the patch analysis value at grid point $n$. Thus the global local patches always have size $2s+1$, but the regional local patches have variable sizes depending on $n$. For $n$ located in the subregion and also far away from the boundaries, the regional local patch has size $2s+1$, while for $n$ near the boundaries ($n+s>n_1$ or $n-s<n_0$), the regional local patch is the intersection, $[n-s,n+s]\cap[n_0,n_1]$, and has a size less than $2s+1$.

For the joint state method, we use the same local patch size, $s=40$. For each grid point $n$ defined either in the global model or in the regional model, we define a global local patch and a regional local patch (where, as before, the regional patch is the intersection, $[n-s,n+s]\cap[n_0,n_1]$, which for some $n=4m$ will be empty). For each such grid point $n$, we define a vector $\mathbf{x}^{(n)}_g$ by taking state values of the global local patch, and $\mathbf{x}^{(n)}_r$ by taking state values of the regional local patch, and we then form a local joint state vector $\mathbf{x}^{(n)}$ by concatenating $\mathbf{x}^{(n)}_g$ and $\mathbf{x}^{(n)}_r$, i.e.,
\begin{equation}
\mathbf{x}^{(n)}=\begin{pmatrix}\mathbf{x}^{(n)}_g\\ \mathbf{x}^{(n)}_r \end{pmatrix}.
\end{equation}
We also form a local observation vector $\mathbf{y}^{(n)}_o$ by taking observations in the local patches (from grid point $n-s$ to $n+s$). We define a local cost function $J^{(n)}(\mathbf{x}^{(n)})$ for grid point $n$ as follows,
\begin{align}
&J^{(n)}(\mathbf{x}^{(n)}) =(\mathbf{x}^{(n)}-\bar{\mathbf{x}}^{(n)}_b)^T(\mathsf{P}^{(n)}_b)^{-1}(\mathbf{x}^{(n)}-\bar{\mathbf{x}}^{(n)}_b)\label{cost function}\\
&+ \left[\mathbf{y}^{(n)}_o-\mathbf{H}^{(n)}(\mathbf{x}^{(n)})\right]^T \mathsf{R}^{-1} \left[\mathbf{y}^{(n)}_o-\mathbf{H}^{(n)}(\mathbf{x}^{(n)})\right]\nonumber\\
&+ \kappa \left[\mathbf{G}^{(n)}_g(\mathbf{x}^{(n)}_g) - \mathbf{G}^{(n)}_r(\mathbf{x}^{(n)}_r)\right]^T \left[\mathbf{G}^{(n)}_g(\mathbf{x}^{(n)}_g) - \mathbf{G}^{(n)}_r(\mathbf{x}^{(n)}_r)\right],\nonumber
\end{align}
where $\bar{\mathbf{x}}^{(n)}_b$ and $\mathsf{P}^{(n)}_b$ are the local mean and the covariance matrix of the background ensemble, respectively, $\mathbf{H}^{(n)}(\mathbf{x}^{(n)})$ is a local observation operator defined as
\begin{equation}
H^{(n)}_{i}(\mathbf{x}^{(n)}) =
\begin{cases}
	(1-\lambda)\,x_{g,j(i)} + \lambda\,x_{r,j(i)}, &\text{if } n_{0} \le j(i) \le n_{1};\\
	x_{g,j(i)}, &\text{otherwise,}
\end{cases}\label{parameter b}
\end{equation}
where $j(i)$ is the observation location of the $i^{th}$ observation in the local patch, $x_{g,j(i)}$ is the global state value at location $j(i)$, and $x_{r,j(i)}$ is the regional state value at location $j(i)$. $\mathbf{G}^{(n)}_g(\mathbf{x}^{(n)}_g)$ is a vector that consists of the state values of the global state at the grid points defined both in the global and the regional local patches. Similarly, $\mathbf{G}^{(n)}_r(\mathbf{x}^{(n)}_r)$ is a vector that consists of the state values of the regional state at the grid points defined both in the global and the regional local patches. $\kappa$ and $\lambda$ are parameters. The third term is a constraint term that penalizes large differences between the estimates of the global and regional model states. We determine the value of x that minimizes the cost function $J^{(n)}(\mathbf{x}^{(n)})$ with the LETKF algorithm \citep{Hunt-et-al-07}.

In general, if our technique were to be applied in an operational setting, the grid points of the global and the regional models within the subregion will not coincide. In that case, to calculate the third term in $J^{(n)}(\mathbf{x}^{(n)})$, an interpolation from the grid points of the regional model to the grid points of the global model or vice versa could be employed before the values of the regional and the global models are subtracted. Similarly, in an operational setting the observations are not at grid points, and $\mathbf{H}^{(n)}$ would then include interpolation.

\section{Model integration}\label{integration}

We define a smoothed regional state for the initial condition of the regional model for integration between analysis times as follows. After the analysis phase, we define spatial transition intervals of length 10 starting from the boundaries and ending inside the subregion. We then modify the regional analysis values in the transition intervals by taking weighted linear averages of the global analysis values and the regional analysis values. We do this in order to make the transition between the global model and the regional model smooth at the boundaries. For $n$ such that $0 \le n < 10$, we modify the regional ensemble members by
\begin{align}
X^r_{k,n_0+n} &\leftarrow (n/10) \, X^r_{k,n_0+n} + (1-n/10) \, X^g_{k,n_0+n},\\
X^r_{k,n_1-n} &\leftarrow (n/10) \, X^r_{k,n_1-n} + (1-n/10) \, X^g_{k,n_1-n},
\end{align}
where $X^g_{k,n}$ and $X^r_{k,n}$ are the values of the $k^{th}$ global and regional ensemble members at grid point $n$, respectively, and the subregion for the regional model is $[n_0,n_1]=[240,720]$.

After performing the above smoothing process, we integrate each global and regional ensemble members for 6 hours using a fourth-order Runge-Kutta method, dividing 6 hours into 24 time steps. We integrate the global ensemble members independent of the regional ensemble members. For the integration of the regional ensemble members, we use the necessary interpolated values of the corresponding global ensemble members outside the subregion at each Runge-Kutta time step to synchronize the global and the regional model at the boundaries.

\section{Results}\label{results}

Before we tested the joint state method and the separate analysis method, we ran forecast cycles with 40 ensemble members using the global and the regional models separately and found that multiplicative covariance inflation factors of 0.024 and 0.02 for the global and the regional analyses, respectively, produce the lowest rms state estimate errors. We henceforth use these values in our data assimilations. For the joint state method, we found that $\lambda=0.9$ and $\kappa=0.04$ in Eqs.~(\ref{cost function}) and (\ref{parameter b}) give the lowest rms state estimate errors, and we use these values in all of our subsequent applications of the joint state method.

\begin{figure}
\includegraphics{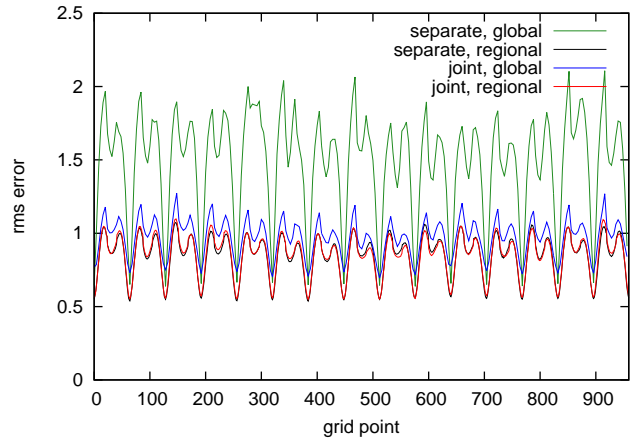}
\caption{Rms errors of the state estimates of the separate analysis and the joint state analysis using the whole region for both the global and the regional models. The rms-error values were averaged over 10000 forecast cycles, discarding the values of 1000 initial cycles. The green and the black colors correspond to the global and the regional values obtained using the separate analysis method. The blue and the red colors correspond to the global and the regional values obtained using the joint state method.}\label{whole region}
\end{figure}

We first tested the separate analysis method and the joint state method without boundaries. That is, we used the whole region for both the global and regional models. Thus, there is no coupling between the global model and the regional model at the boundaries during the integration phases. In this setup, aside from the correlations induced by common observations in their assimilations, the separate analysis method corresponds to having independent global and regional forecasts. For the joint state method, the coupling between the global and regional models occurs only at the analysis phases. Figure~\ref{whole region} shows the rms errors of state estimates given by the means of the ensemble members as a function of the grid point.
The values were averaged over 10000 forecast cycles, discarding the values of 1000 initial cycles. The green and the black colors correspond to the global and the regional values obtained from the separate analysis method. The blue and the red colors correspond to the global and the regional values obtained from the joint state method. Error minima occur at the observation points. The figure shows that the two regional rms errors are almost the same, while the global rms errors from the joint state method are much lower than the global rms errors of the separate analysis case indicating that, as one would expect, the information from the regional model substantially improved the estimate of the global model.

\begin{figure}
\includegraphics{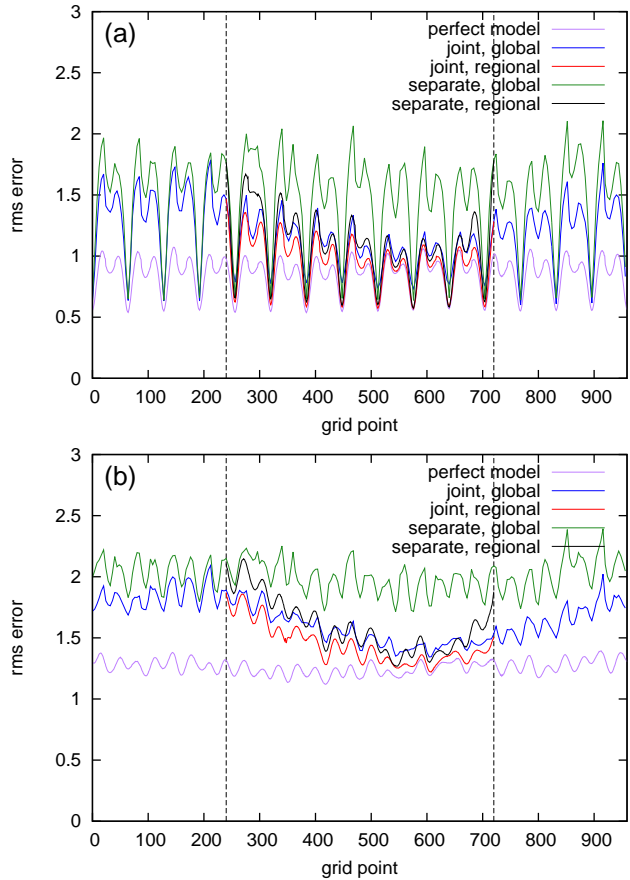}
\caption{Rms errors of (a) the state estimates (b) 1 day forecasts of the separate analysis and the joint state analysis. The color scheme is the same as in Fig.~\ref{whole region}. The additional purple curves show the rms-error values when assimilations were done globally using the true model (Lorenz model 3). The two vertical dashed lines at grid points 240 and 720 indicate the boundaries of the subregion.}\label{subregion}
\end{figure}

Now, we take a subregion $[n_0,n_1]=[240, 720]$, and introduce coupling between the global model and the regional model at the two boundaries during the integration phase. Figures~\ref{subregion}(a) and \ref{subregion}(b) show the rms errors of the analysis and of a 1 day forecast, respectively, using the same color scheme as in Fig.~\ref{whole region}.
The two vertical dashed lines at grid points 240 and 720 indicate the boundaries of the subregion. The additional purple curves show the rms-error values in the perfect model scenario in which the forecast model was the true model (Lorenz model 3) which was used globally throughout the entire space. We view this as setting a standard for the best that could ever be done. These figures show that the joint state method performs better than the separate analysis method for both the global prediction and the regional prediction. We note that the global forecast obtained from the joint state method is better than the corresponding one from the separate analysis method even outside the subregion. This can be explained by the fact that the better global state estimates inside the subregion at the analysis phases can make better forecasts outside the subregion during the integration phases, and these better forecasts outside the subregion can make the regional forecasts better inside the subregion by providing better information at the boundaries during the integration phases. We also note that the global analysis improvements that result from use of the joint state method are greater to the right of the subregion than to its left. This is consistent with the fact \citep{Lorenz-05, Yoon-et-al-10} that, for these models, waves (and hence the information they carry) have group velocities that are predominantly rightward.

\section{Discussion and conclusion}\label{conclusion}

In this paper we formulated a joint state method for regional forecasting. Using simulations employing simple models, we have numerically tested our method by comparing analysis and forecast results obtained using our method with results obtained using a separate analysis method. We found that the global forecast in the whole region and the regional forecast in the subregion are both noticeably improved when the joint state method is used compared to when the separate analysis method is used.

This work suggests several topics for future work. Most importantly, will the encouraging results from experiments using our Lorenz model set-up continue to apply when tests on real situations are done? What is the effect of regional model error? What are the benefits of applying our coupled analysis scheme to situations with multiple (perhaps overlapping) regional analyses?

\section{Acknowledgments}

This work was supported by NSF grant ATM-0935538 and ONR grant N000140910589.


\end{document}